# Assessing Sensitivity of Brain-to-Scalp Blood Flows in Laser Speckle Imaging by Occluding the Superficial Temporal Artery

Yu Xi Huang,[1,†] Simon Mahler,[1,†,*] Maya Dickson,[1,†] Aidin Abedi,[2,3,4] Yu Tung Lo,[2,5] Patrick D. Lyden[6], Jonathan Russin,[2,3] Charles Liu,[2,3,**,‡] Changhuei Yang[1,‡]

**Affiliations**

[1]Department of Electrical Engineering, California Institute of Technology; Pasadena, CA 91125, USA.
[2]USC Neurorestoration Center, Department of Neurological Surgery, Keck School of Medicine, University of Southern California; Los Angeles, CA 90033, USA.
[3]Rancho Research Institute, Rancho Los Amigos National Rehabilitation Center; Downey, CA 90242, USA.
[4]Department of Urology, University of Toledo College of Medicine and Life Sciences; Toledo, OH 43614, USA.
[5]Department of Neurosurgery, National Neuroscience Institute, Singapore 308433.
[6]Department of Physiology and Neuroscience, Zilkha Neurogenetic Institute, and Department of Neurology, Keck School of Medicine, University of Southern California, Los Angeles, CA 90033 USA.

[†]These authors contributed equally to this work.
[‡]These authors co-supervised this work.
**Corresponding authors:**
[*]Emails: mahler@caltech.edu, sim.mahler@gmail.com
[**]Email: cliu@usc.edu

**Teaser**

We combined a multi-channel laser system with temporal occlusion to experimentally distinguish brain and scalp signals non-invasively.

**Abstract**

Cerebral blood flow is a critical metric for cerebrovascular monitoring, with applications in stroke detection, brain injury evaluation, aging, and neurological disorders. Non-invasively measuring cerebral blood dynamics is challenging due to the scalp and skull, which obstruct direct brain access and contain their own blood dynamics that must be isolated. We developed an aggregated seven-channel speckle contrast optical spectroscopy system to measure blood flow and blood volume non-invasively. Each channel, with distinct source-to-detector distance, targeted different depths to detect scalp and brain blood dynamics separately. By briefly occluding the superficial temporal artery, which supplies blood only to the scalp, we isolated surface blood dynamics from brain signals. Results on 20 subjects show that scalp-sensitive channels experienced significant reductions in blood dynamics during occlusion, while brain-sensitive channels experienced minimal changes. This provides experimental evidence of brain-to-scalp sensitivity in optical measurements, highlighting optimal configuration for preferentially probing brain signals non-invasively.

**Introduction**

Monitoring cerebral blood flow dynamics is crucial for understanding brain health, diagnosing neurological conditions such as stroke and structural brain injury, and developing effective treatments *(1–3)*. While spatial imaging of the brain has significantly advanced in the last decades with modalities such as magnetic resonance imaging (MRI) *(4)*, computed tomography (CT) *(5)*, and X-rays *(6)*, providing high-resolution images of the brain, temporal imaging—particularly for measuring cerebral blood dynamics—remains challenging with these high spatial resolution modalities *(7)*.

To effectively monitor cerebral blood flow dynamics, a temporal resolution of 20 Hz or higher is desired. This is based on the typical resting heart rate in adults, which ranges between 50 and 100 beats per minute (bpm) *(8)*. Each heartbeat follows a cardiac cycle comprising distinct phases, such as the P-wave, QRS complex, ST segment, and T-wave, as observed in an electrocardiogram *(9–11)*. Some of these phases are separated by less than one-tenth of the cardiac cycle duration *(12–14)*. Monitoring blood flow dynamics effectively implies detecting each phase of the cardiac cycle, necessitating a temporal resolution of ten times the heart rate. Considering a nominal heart rate of 60 bpm (1 Hz), the cardiac cycle would include temporal features that we would like to sample at 20 Hz or more. This sets the device's sampling rate required to accurately capture the temporal blood dynamics.

For the past few decades, characterizing cerebral perfusion capacity such as cerebral blood flow has been performed by using various nuclear imaging techniques *(15–17)*. However, these methods have not seen widespread adoption due to technical inefficiencies, logistical challenges, and high costs. Neuroimaging techniques such as positron emission tomography, single photon emission computed tomography, and perfusion computed tomography can provide insights into cerebral blood flow and brain perfusion *(15–17)*, but they are limited by temporal resolution, side effects, logistical constraints, and the inability to assess physiological stimuli effectively. Functional magnetic resonant imaging (fMRI) is a commonly used research tool that offers comprehensive assessments of cerebral blood flow, blood volume, and blood oxygenation *(18–20)*. However, its temporal resolution is limited to few images per second, leaving a gap in real-time blood dynamics monitoring. In addition, the high cost and operational complexity of MRI make it unsuitable for rapid screening.

All the above-mentioned nuclear imaging methods lack sufficient temporal resolution for measurements of effective cerebral blood dynamics. Electroencephalogram (EEG), which is a non-imaging technique, does have the temporal resolution and depth of penetration for brain dynamics monitoring *(12, 14, 21, 22)*. EEG, however, primarily captures electrical impulses produced by brain cells and measures the brain's activity in the form of electrical waves and is not capable of measuring blood flow dynamics directly. Another possible candidate is transcranial Doppler ultrasound (TCD) *(23)*. TCD can measure cerebral blood flow with high temporal resolution (typically 50-100 Hz). However, TCD is not possible in every patient due to limited skull penetration of sound waves *(24–26)*. Also, skull bone shows a much higher acoustic impedance (resistance to sound wave transmission) compared to soft tissues, due to the skull's air gaps. As such, TCD is mostly used on the temporal area of the head *(26)*. TCD is also limited to measuring large vessels' cerebral blood flow, lacking the ability to assess microvascular dynamics *(25, 27)*.

For monitoring cerebral blood dynamics, optical transcranial measurement modalities offer compelling advantages – non-ionizing radiation, cost-effective equipment and ease of use. Through diffuse correlation spectroscopy (DCS) or speckle contrast optical spectroscopy (SCOS) *(25, 27–33)* (also named speckle visibility spectroscopy (SVS) *(34–39)*), brain blood characteristics, such as cerebral blood volume (CBV) through optical signal attenuation, and cerebral blood flow (CBF), can be measured with high temporal resolution (40-100 Hz). Among optical imaging methods, SCOS has recently gained prominence in monitoring blood flow and blood volume *(27, 29, 40)*. In contrast to DCS, SCOS measures cerebral blood dynamics using affordable consumer cameras *(25, 28, 29, 41)* by capturing fluctuating speckle patterns. Unlike transcranial Doppler, which measures blood flow in major vessels, SCOS is sensitive to all blood cell movements, allowing the quantification of microvascular hemodynamics, critical for assessing ischemic vascular pathologies *(25, 31)*. Lastly, the new compact SCOS design *(25, 28, 31, 41)*, which mounts directly on the subject's head, removes the need for bulky optical and electrical components, thus increasing portability and reducing motion artifacts in the measurements.

Despite its potential, demonstrating that SCOS and other optical imaging techniques can effectively probe brain signals over scalp and skull layers was an unresolved matter. While numerous numerical studies have assessed the depth sensitivity of optical methods *(42–44)*, to the best of our knowledge, no experimental validation of the depth sensitivity has yet been achieved.

To address this gap, we present our experimental investigations to simultaneously monitor blood flows in the scalp, skull/meninges, and brain, with the goal of evaluating the influence of scalp and skull layers on brain blood dynamics measurements for optical techniques. For this purpose, we developed a seven-channel cerebral blood flow and volume monitoring SCOS system. Each detecting channel was strategically positioned at distinct distances relative to the laser source, enabling selective measurement of scalp or brain blood flow. To isolate scalp blood flow from brain blood flow, we performed temporary occlusion on the superficial temporal artery (STA), which supplies blood to the scalp and skull layers but not the brain. Our results demonstrate that channels primarily measuring scalp flow exhibited significant decreases in blood flow during the temporal occlusion, whereas channels with higher brain specificity exhibited minimal decreases. We also compared experimentally the difference in sensitivity between flow and volume measurements. This study provides the first experimental evidence of varying brain-and-scalp sensitivity in optical measurements, providing crucial insights into the optimal device configuration for preferentially probing brain signal over scalp signal.

## Results

To investigate the effects of superficial temporal artery occlusion on scalp and brain blood flow, we developed a seven-channel SCOS system (Fig. 1) to simultaneously monitor blood dynamics in the scalp, skull, and brain, assessing the influence of superficial layers on cerebral measurements. The device features a 3D-printed head mount designed for secure placement over the frontal temporal region, targeting the superficial temporal artery. It incorporates a single illumination fiber shining laser light at 830 nm and seven 600 μm core diameter detection fibers positioned at varying source-detector (S-D) distances to capture blood flow dynamics across layers. The detecting fibers are bundled and imaged onto a scientific CMOS camera for simultaneous signal acquisition. The laser intensity adheres to American National Standards Institute (ANSI) safety limit for skin exposure to an 830 nm laser beam (3.63 mW/mm²) *(45)*. The mount's design accommodates head curvature, minimizing movement artifacts. Velcro straps secure the device without restricting scalp blood flow. See Materials and Methods for more details about the device's experimental arrangement and operating principles. Temporal artery occlusion was performed by gently applying pressure near the ear bone, isolating scalp blood flow while preserving cerebral circulation, providing a minimally invasive and repeatable method for all subjects.

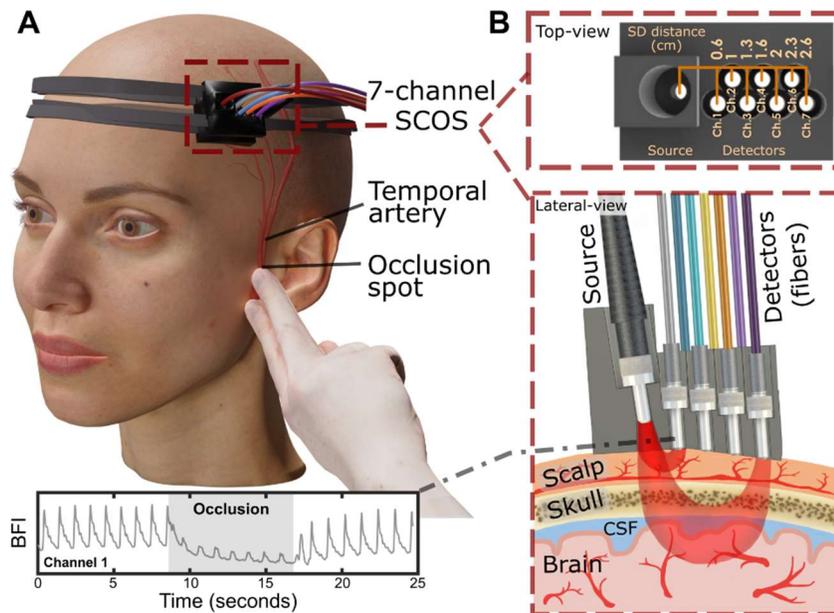

**Fig. 1. Experimental arrangement of the SCOS system for measuring cerebral blood dynamics during superficial temporal artery (STA) occlusion.** (**A**) 3D visualization of the SCOS device positioned over the temple region and the occlusion site near the ear bone. (**B**) Top and lateral views of the device, illustrating different detecting channels for sensing the scalp, skull, and brain layers.

In Fig. 1, we present a typical CBF time trace from Channel 1 (S-$D_{Ch1}$ = 0.6 cm) with the superficial temporal artery occluded between the 8$^{th}$ and 16$^{th}$ seconds. In all our results, the CBF time traces are presented in blood flow index (BFI) units, and the CBV time traces in blood volume index (BVI) units, both of which were normalized to display relative changes. As shown in Fig. 1A, the significant dip in the CBF signal during the occlusion period indicated a substantial reduction in scalp blood flow. The different channels in Fig. 1B are located at different S-D distance, i.e. at different distances between the illumination point on the head (source) and the detection point (detector) where the scattered light is collected back. As shown in Fig. 1B, this dip should become less pronounced with increasing channel numbers and corresponding S-D distances due to decreasing sensitivity to scalp blood flow. The depth of light penetration into the head correlates with the S-D distance, with channels at larger S-D distances probing deeper structures *(35)*. Based on numerical studies *(42–44)* and prior experimental work with interferometric SCOS *(35)*, we anticipate Channel 1 (S-$D_{Ch1}$ = 0.6 cm) to primarily probe scalp blood flow, Channels 2 (S-$D_{Ch2}$ =

1.0 cm) and 3 (S-D$_{Ch3}$ = 1.3 cm) to primarily probe scalp and skull blood flow, Channel 4 (S-D$_{Ch4}$ = 1.6 cm) to probe scalp, skull, and some brain blood flow, and substantial gains in brain signal sensitivity from Channels 5 to 7 (2.0, 2.3, and 2.6 cm).

Figure 2 displays the CBF results from the seven channels, with Channel 1 (0.6 cm) at the top and Channel 7 (2.6 cm) at the bottom of the figure. Figure 2A shows the results when a temporal occlusion was applied between the 8$^{th}$ and 16$^{th}$ seconds as in Fig. 1A, while Fig. 2B shows the control results, where a non-occlusive press on the ear was performed during the same time frame. This mimicked the movement induced by the occlusion press, ensuring that the observed results were not attributed to motion artifacts. The recordings of Figs. 2A and 2B were conducted less than 5 minutes apart.

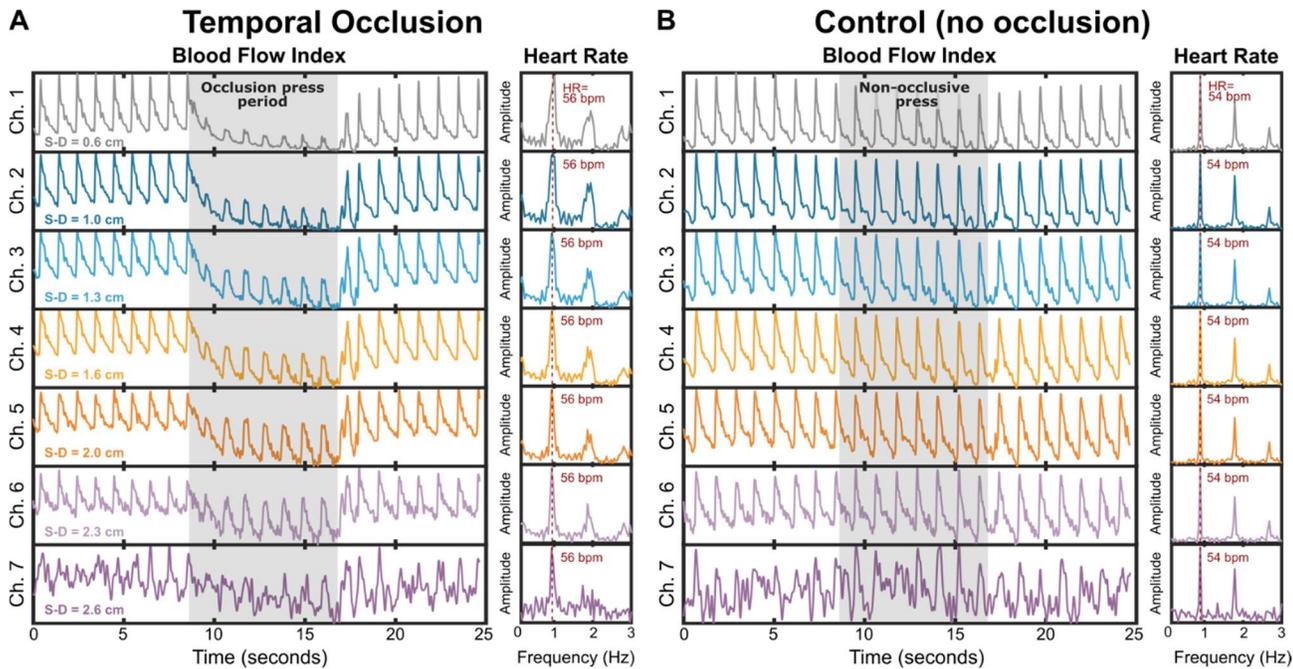

**Fig. 2. Typical cerebral blood flow (CBF) time traces during superficial temporal artery occlusion and non-occlusive pressure.** (**A**) CBF traces from the seven channels during superficial temporal artery occlusion (8$^{th}$–16$^{th}$ seconds). Channels targeting the scalp show a significant dip in CBF, while brain-targeting channels exhibit a smaller change. (**B**) CBF traces under non-occlusive pressure applied to the ear instead of the superficial temporal artery. No significant differences in CBF dynamics are observed across channels, confirming the specificity of the superficial temporal artery occlusion in blocking scalp blood flow.

As shown, the seven channels were synchronized for both cases with minimal delays, as expected due to the close proximity of the channels' positions, exhibiting identical oscillation frequencies corresponding to the heart rate and cardiac cycle. The heart rate was determined from the frequency graphs (right panels), obtained by applying a Fourier transform to the CBF time traces *(28, 35, 38)*. Note that the frequency graphs in Fig. 2A exhibit a broader heart rate peak than those in Fig. 2B due to the change of CBF waveforms during temporal occlusion. A significant dip in the CBF can be observed during the occlusive press, while there was no change in the CBF time trace during the non-occlusive press. This validates the effectiveness of the occlusive press in reducing CBF.

As the channel number and corresponding S-D distance increased, the dip in CBF during the occlusion became progressively smaller, Fig. 2A. Additionally, the amplitude of the cardiac waveforms significantly decreased during the occlusion period for Channel 1 (0.6 cm) and increased back to baseline with greater S-D distances. By Channel 7 (2.6 cm), the dip and change in waveform during occlusion were nearly absent. Despite this, the heart rate frequency graph indicates that Channel 7 was still able to capture accurate CBF signals. These findings demonstrate that the CBF time trace of Channel 1 was heavily influenced by the temporal occlusion, with its impact diminished as the S-D distance increased. This aligns with our expectation that the CBF dip becomes less pronounced with increasing channel numbers, reflecting reduced sensitivity to scalp blood flow. In Fig. 2B, no changes were observed in the CBF time trace during the non-occlusive press, and all channels were highly synchronized and exhibited the same heart rate. These indicate that the motion artifacts induced by the non-occlusive press had no effect on the recorded CBF. This also indicates that blood flow from the scalp and brain layers cannot be directly distinguished without an intervention specifically targeting one of the two layers.

Next, we repeated the measurement of Fig. 2 across 20 subjects and quantified the extent of CBF signal reduction during temporal occlusion for each channel (Fig. 3). See Materials and Methods for details about the subject cohort and study protocol. First, the CBF signal was segmented into three periods (a, b, and c), Fig. 3A:

a. **Pre-occlusion**: Segment from the start to the 8$^{th}$ second.
b. **During occlusion**: Segment from the 8$^{th}$ to the 16$^{th}$ second.
c. **Post-occlusion**: Segment from the 16$^{th}$ second to the end.

For each segment, we calculated the mean value $<BFI>$ and the standard deviation $\sigma$ of the blood flow index. Figure 3B shows the mean ratio of the post-occlusion to pre-occlusion segment $<BFI_c>/<BFI_a>$ (amber curve with diamonds) and the mean ratio of the during-occlusion to pre-occlusion segment $<BFI_b>/<BFI_a>$ (blue curve with crosses) as a function of the S-D distance for each of the seven channels across the 20 subjects. The dashed lines represent the results from the control case (i.e., non-occlusive press). The gray solid line represents a ratio of one. Error bars indicate the statistical standard deviation across the 20 subjects. As shown, the mean ratio for the post-occlusion to pre-occlusion is similar to the ratios observed in the control case, i.e. approximately one, suggesting little to no change in CBF and overall blood flow index before and after the occlusion. However, the mean ratio for the during-occlusion to pre-occlusion segment is significantly lower at an S-D distance of 0.6 cm, around 0.65, and increases back to 1 with increasing S-D distance. At an S-D distance of approximately 2 cm, the mean ratio converges to the control case, indicating that an S-D distance of at least 2.0 cm is necessary to minimize the influence of scalp blood flow. See Supplementary Material (Fig. S5) for individual mean ratio graphs for each of the 20 subjects.

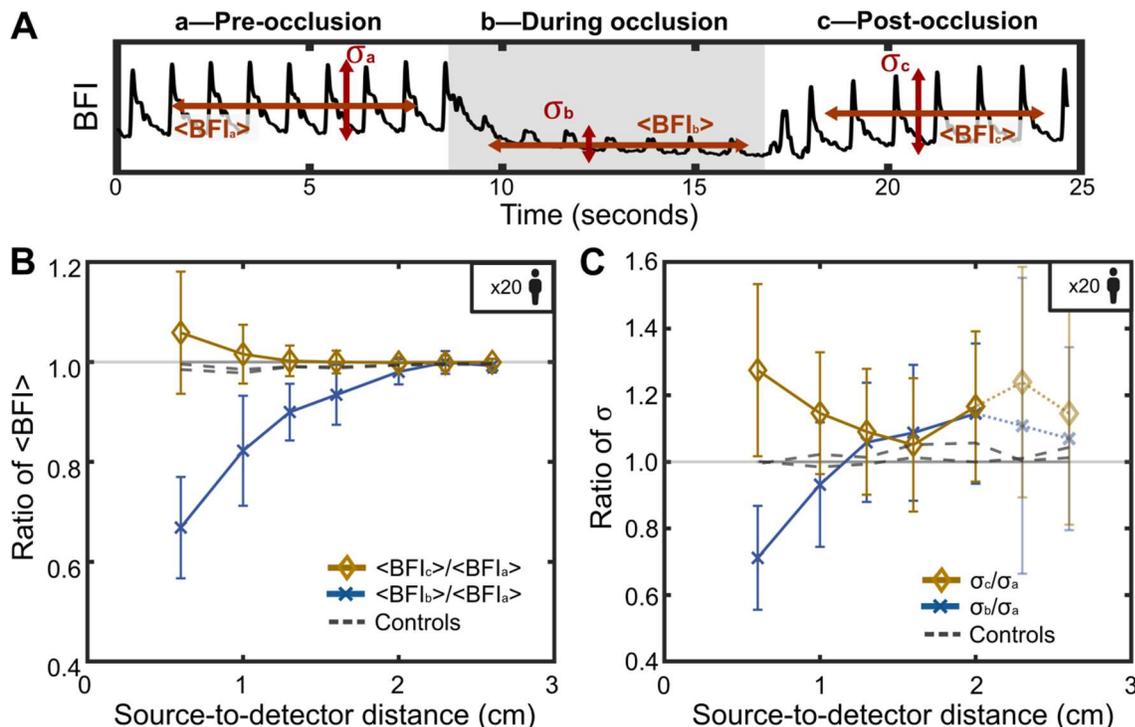

**Fig. 3. Effects of the superficial temporal artery occlusion on the scalp, skull, and brain blood flows measured from 20 subjects.** (A) Segmentation of the blood flow time trace into three sections: pre-occlusion, during occlusion, and post-occlusion. Quantification of the blood flow mean <BFI> and blood flow variance σ. (B) Average ratio of the blood flow mean among the three sections of **A**. (C) Average ratio of the blood flow variance among the three sections of **A**.

Figure 3C displays the standard deviation ratios. For Channels 6 (2.3 cm) and 7 (2.6 cm), the results are faded out due to insufficient SNR in some subjects, as reflected by the large error bars. The standard deviation ratio for the post-occlusion to pre-occlusion segments remains constant and approaches the control case, similar to the behavior shown in Fig. 3B. In contrast, the ratio for the during-occlusion to pre-occlusion segment starts around 0.7 and gradually approaches the control case as the S-D distance increases. The convergence occurs at an S-D distance of approximately 1.3 cm, notably earlier than the S-D distance of 2.0 cm in Fig. 3B. We attribute this difference to the fact that Fig. 3B and 3C quantify different metrics.

Figure 3B illustrates the mean value of blood flow index (BFI), which quantifies the average blood flow level during each segmented period. This accounts for mean blood circulation over a segmented time period of 8 seconds, encompassing multiple cardiac cycles. During an occlusion, the total amount of flowing blood decreases in the scalp, resulting in a corresponding reduction in the mean BFI. Channels with shorter S-D distances exhibited a larger decrease in $<BFI>$ during an occlusion, while channels with longer S-D distances exhibited a smaller decrease due to their lower sensitivity in scalp blood flow. Note that even at very large S-D distances (e.g., greater than 2.6 cm), light would still pass through the scalp and skull layers (Fig. 1B). Consequently, a small dip in the CBF signal was expected during an occlusion, showing up as a slightly reduced $<BFI>$. This explains why the ratio of mean BFI during to pre-occlusion ($<BFI_b>/<BFI_a>$) in Fig. 3B remained slightly below one, even at larger S-D distances, as the occlusion always impacts the overall CBF value.

Notably, the ratio of <BFI> post- to pre-occlusion ($<BFI_c>/<BFI_a>$) exceeded one at short S-D distances due to the sharp increase in blood flow after the occlusion is released. This sharp rise occurred because the blockage temporarily restricted blood, and when released, blood flow surged to levels higher than pre-occlusion in some subjects. As the S-D distance increased, the ratio of $<BFI>$ post- to pre-occlusion gradually converged to one, due to lower sensitivity in scalp blood flow. For control cases (e.g., Fig. 2B), the ratio of $<BFI>$ remained consistently around one, as there was no significant change in the CBF signal during the non-occlusive press. Thus, the ratio of $<BFI>$ may serve as an effective controlled experimental metric for estimating brain and scalp sensitivity of any optical transcranial devices. Similar approaches have been employed in previous studies to characterize cerebrovascular reactivity via CBF variations during breath-holding *(25, 27)*.

Figure 3C illustrates the standard deviation of the blood flow index (BFI), which was influenced by the amplitude of the cardiac cycle, occurring on a faster timescale—typically within a second or less. The standard deviation of BFI primarily quantified the variation in blood flow between the peak pressure (systolic phase peak) and the lowest pressure (diastolic phase endpoint) over a series of cardiac cycles. A greater difference in blood flow during a cardiac cycle—corresponding to a higher blood pressure variation between the systolic peak and diastolic endpoint—resulted in a higher standard deviation of the BFI. During an occlusion, the scalp blood flow waveform was significantly reduced due to the decreased level of blood circulation. Additionally, the repeatability of the waveform (i.e., the cardiac cycle) was markedly diminished because of the blood blockage. As a result, the difference between the systolic phase peak and the diastolic phase endpoint in scalp blood flow was observed to be smaller, leading to a reduced standard deviation of the BFI. Channels with shorter S-D distances showed a larger decrease in standard deviation during an occlusion, again due to their high sensitivity to scalp blood flow, while channels with longer S-D distances exhibited less or no decrease due to their lower sensitivity to scalp blood flow. In this case, while the difference between the systolic phase peak and diastolic phase endpoint in scalp blood flow remained small, the corresponding difference in brain blood flow remained large, as the occlusion did not affect brain blood flow. Consequently, the cardiac cycle was dominated by brain blood flow, and the standard deviation of the BFI was unaffected by the occlusion.

The ratio of post- to pre-occlusion ($\sigma_c/\sigma_a$) exceeded one at short S-D distances due to the sharp increase in blood flow after the occlusion was released, with the blood flow surging at a higher peak pressure than that observed pre-occlusion. As the S-D distance increases, this ratio remained slightly above one. For control cases (Fig. 2B), the $\sigma$ ratio remained consistently around one, as there was no significant change in the CBF signal during the non-occlusive press. Due to the differences between Fig. 3A and Fig. 3B—where Fig. 3A measured more the overall blood levels and Fig. 3B captured blood flow differences between systole and diastole—the convergence to the control case varied between the two. Specifically, the results in Fig. 3A converged to the control case at a larger S-D distance.

Next, we quantified the reduction in CBF and CBV waveforms during the occlusion. The results are shown in Fig. 4. In Fig. 4A, we segmented both the CBF and CBV signals into the same three segments (a, b, and c) as in Fig. 3A. For each segment, we calculated the average cardiac cycle waveform *(27)*. The results are presented in Fig. 4B for CBF by calculating the blood flow index (BFI) waveform amplitude and Fig.

4C for CBV by calculating the blood volume index (BVI) waveform amplitude. Note that Figs. 4A to 4C show the CBF and CBV data for Channel 1 (0.6 cm) of the subject featured in Fig. 2.

As shown, the CBF waveform underwent notable changes during the occlusion, particularly in the dicrotic notch and peak pressure, while the CBV waveform remained largely unchanged. However, the overall values of the blood flow index (BFI) and blood volume index (BVI) significantly decreased during the occlusion period. To further characterize these changes, we calculated the peak amplitude $I$ for both CBF and CBV in each of the three segments, Figs. 4B and 4C. The results, shown in Fig. 4D for CBF and Fig. 4E for CBV, exhibited similar trends but with differing rates of convergence. While the amplitude ratio during to pre-occlusion was less than one at shorter S-D distances, it approached a ratio of one around a S-D distance of 1.3 cm for CBF and a S-D of 2.5 cm for CBV. This disparity highlights the difference in brain sensitivity between CBF and CBV measurements in SCOS and optical systems.

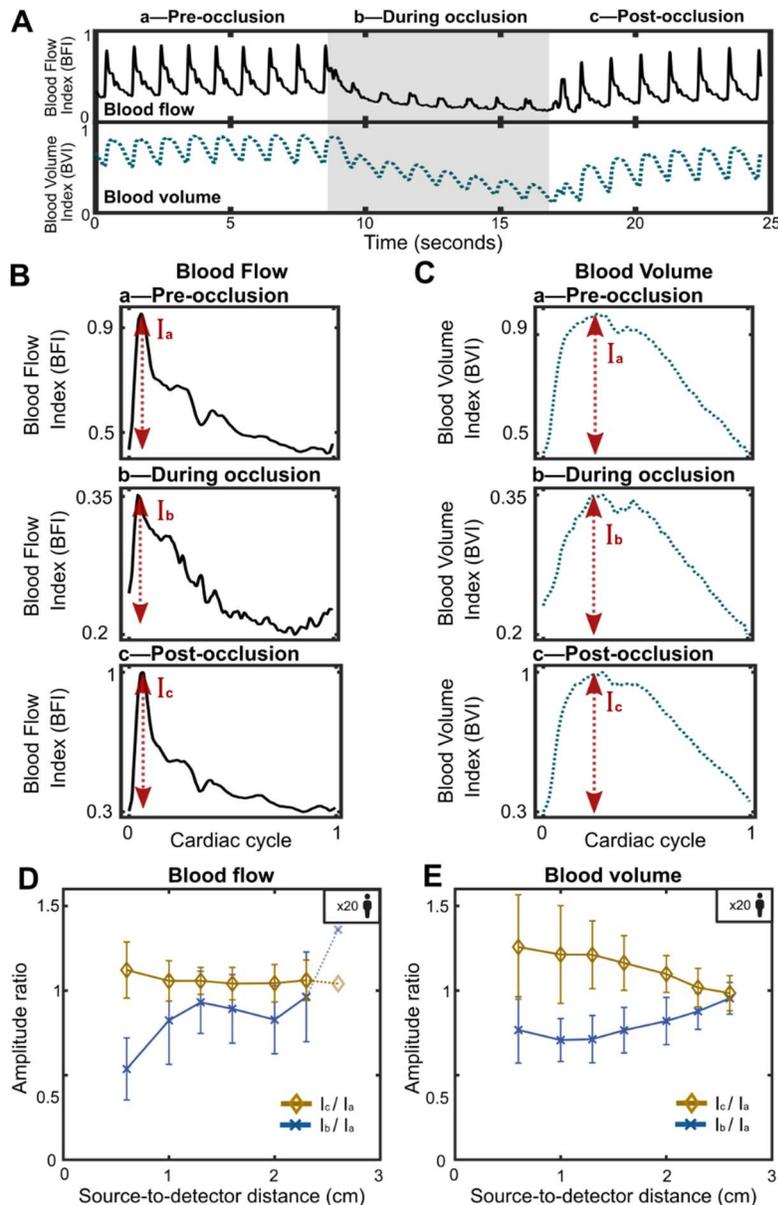

**Fig. 4. Effects of superficial temporal artery occlusion on CBF and CBV.** (**A**) Representative time traces of Channel-1 (S-D=0.6 cm) CBF and CBV time traces during superficial temporal artery occlusion. (**B**) CBF and (**C**) CBV cardiac pulse waveforms corresponding to the segment periods defined in **A**. Amplitude ratio of (**D**) CBF and (**E**) CBV as a function of the source-to-detector distance. The results show that with increasing S-D distance, CBF is less influenced by the temporal occlusion effects (i.e. less sensitive to scalp blood flow) compared to CBV. These experimental findings align with numerical simulations presented in Fig. 1 of *(43)*.

This experimental result aligns well with previous numerical simulation results reported in *(43)*, where the brain-to-scalp sensitivities of optical BFI and optical absorption (as measured in BVI) were simulated using a double-layer model. Their findings, shown in Fig. 1 of *(43)*, demonstrated that BFI is intrinsically more brain-specific than absorption (BVI), achieving higher brain-to-scalp sensitivity at a given S-D separation. For example, in *(43)*, at an S-D distance of 1.3 cm, BFI achieved a brain-to-scalp sensitivity of 15%, while BVI required an S-D distance of over 2.5 cm to achieve the same sensitivity. These findings are consistent with our experimental observations in Figs. 4D and 4E, where a lower scalp sensitivity during temporal occlusion was observed at an S-D distance of 1.3 cm for CBF and at 2.5 cm for CBV. This can be attributed to the fact that blood flow (BFI) in gray matter is approximately six times greater than in the scalp, whereas hemoglobin concentration—which relate to blood volume (BVI)—is only two to three times greater in gray matter compared to the scalp *(46)*. These results highlight the superior brain specificity of CBF measurements obtained using speckle contrast or diffuse correlation systems at similar S-D distances when compared to CBV measurements from absorption-based systems and modalities.

We further analyzed the periodic curve patterns illustrated in Fig. 4B, comparing the autocorrelation of cardiac periods across blood flow data before, during, and after temporal occlusion. Despite thorough examination, no consistent changes in curve shapes were observed. This inconsistency may stem from slight variations in the cardiovascular physiological layout around the temporal region. For additional details, refer to Supplementary Material Figs. S6 and S7, which includes an example confusion matrix and autocorrelation trends for both CBV and CBF curves.

## Discussion

We present experimental data quantifying the sensitivity of scalp and brain blood flows in SCOS and other optical methods as a function of source-detector (S-D) distance, alongside a direct comparison of flow and volume sensitivity. This study employed a brief occlusion of the superficial temporal artery during CBF and CBV recordings, showcasing a robust method to disentangle layers of blood flow dynamics and isolate scalp-specific signals through a safe and non-invasive approach.

Our findings represent a significant advancement in non-invasive cerebrovascular monitoring, offering two pivotal contributions. First, we establish an experimental framework for estimating brain to scalp sensitivity in optical systems. Second, we introduce temporal artery occlusion as a simple yet effective intervention for isolating brain-specific signals from disruptive scalp dynamics. This approach holds promise for cerebrovascular monitoring, particularly in investigating how headaches influence scalp and brain blood flow, which remains an open question in the field.

Our results, derived from a cohort of 20 subjects, demonstrate that CBF exhibits relatively low sensitivity to scalp blood dynamics at S-D distances greater than 1.5 cm, while CBV requires a significantly larger S-D distance of 2.5 cm to achieve the same sensitivity, in agreement with prior numerical studies *(43, 46)* This finding underscores the importance of incorporating CBF measurements besides CBV for cerebrovascular monitoring and highlight the utility of SCOS and DCS devices, which are capable of simultaneously detecting both CBF and CBV. Additionally, our results identify an optimal S-D distance of at least 2.0 cm to minimize scalp sensitivity. However, brain sensitivity continues to improve beyond this distance, establishing 2.0 cm as the lower bound for brain blood flow measurements. While the mathematical definition of SCOS has been established for two decades *(30, 39)*, its experimental implementation for cerebral blood flow monitoring is relatively new, with pioneering works in 2018 *(32)* and 2020 *(36)*. Since then, several SCOS systems have been developed by independent research groups to effectively measure CBF and CBV. Notable recent SCOS systems include Openwater bedside SCOS system for stroke detection *(25, 31)*, fiber-based SCOS for mental task probing *(29)* and brain injury in mice *(47)*, compact-SCOS for stroke-risk prediction *(27, 28)* and brain injury investigation *(41)*, and interferometric SCOS (also named iSVS) for cerebral blood flow measurement on human and animals such as rabbits *(35, 36)*.

This variety of systems and applications underscores the reproducibility and effectiveness of SCOS while emphasizing the need for an experimental method to differentiate scalp and brain sensitivities—a gap this work aims to address. Notably, all the SCOS systems referenced utilized a source-to-detector (S-D) distance of 3.0 cm or greater, highlighting the critical importance of carefully selecting the S-D distance in optical systems to optimize sensitivity and accuracy.

Despite the promising results, several limitations must be acknowledged. First, the low SNR in deeper channels, particularly Channels 6 and 7 (2.3 and 2.6 cm), presents challenges in reliably interpreting brain-specific signals at greater S-D distances. Measuring CBF with relatively high SNR is more challenging than CBV at the same S-D distance. To address this, we plan to combine our previous experience in compact SCOS as in Ref *(28)* and develop a custom-designed extended camera spanning over 3 cm in one dimension. This setup will enable simultaneous and continuous measurement of blood dynamics across S-D distances ranging from 0.5 to 3.5 cm, improving spatial resolution and data quality. Second, the superficial temporal artery occlusion method, while effective in isolating scalp signals, relies on precise manual placement and consistent pressure application, which may introduce variability. A mechanically designed system with pre-applied, controlled pressure could improve consistency for the same subject but may still face challenges in maintaining uniformity across different individuals. Third, the study's relatively small sample size of 20 subjects may limit broader generalization. We believe, however, that increasing the cohort size would not significantly enhance the statistical results, as individual variations in scalp and skull thickness inherently contribute to statistical deviations. Finally, further exploration of waveform dynamics, such as the distinct behaviors of CBF and CBV during occlusion, could yield deeper insights into the physiological responses of cerebral and scalp vasculatures under varying conditions.

## Materials and Methods

### Optical methods for measuring cerebral blood flow

Transcranial optical measurement is an appealing technique due to its non-ionizing nature and portability. It enables the quantification of brain blood metrics, such as cerebral blood volume (CBV) via optical signal attenuation *(29, 41)*, and cerebral blood flow (CBF) through diffuse correlation spectroscopy (DCS) *(36, 40, 48–51)* or speckle contrast optical spectroscopy (SCOS), also known as speckle visibility spectroscopy *(34–39)*.

Near-infrared light with a wavelength of 785 nm or higher is typically employed in optical CBF and CBV measurements, due to its ability to penetrate effectively through both the scalp and skull with lower scattering coefficient *(52)*. By transmitting infrared light through one point (source) on the skull and detecting it at another (detector), CBV can be determined by measuring light attenuation *(25, 27, 29)*. If a coherent light, like a laser, is used, CBF can also be measured by tracking fluctuations in the laser speckle pattern, where the blood movement within the brain causes these speckles to fluctuate—with faster blood flow leading to faster fluctuations.

In DCS *(36, 40, 48–51)*, a single photon avalanche diode (SPAD) or a photodetector is used to collect the light. Typically, a single speckle or a small collection of speckles is temporally detected, where a DCS algorithm calculates the signal's time correlation to extract the dynamics. DCS generally employs a fast (MHz range) detector with high quantum efficiency and low detector noise characteristics. For more information about transcranial optical methods, we recommend the following references *(49–51, 53–55)*.

### Speckle contrast optical spectroscopy (SCOS)

The alternate approach, SCOS, is based on the use of spatial ensemble and is an off shoot of laser speckle contrast imaging (LSCI), which typically uses laser speckles for visualizing blood vessels in diverse types of living tissues *(38, 55–58)*. When a coherent laser beam is directed onto the sample, light will experience multiple random scattering events before exiting the sample, causing a granular light in appearance, called speckles. These speckles arise from the mutual interference of light traveling along different trajectories and occur on rapid timescales *(55, 58–62)*. As components (such as red blood cells) within the sample move, the speckle field undergoes dynamic changes related to the inherent dynamics of the sample *(55, 58, 59)*. In SCOS, the speckle fields are temporally recorded by using a camera with a large number of pixels (millions) and a high-frame rate (above 30 FPS). To achieve this, the camera exposure time is set longer than the speckle decorrelation time (typical exposure time ranges between 1 to 10 ms for blood flow imaging), ensuring that the recorded speckle patterns represent an ensemble average of the decorrelation events. As the camera captures multiple images (typically 30 to 100 images per second), the dynamic changes in the speckle field can be quantified by calculating the degree of blurring, or more specifically, the speckle contrast value $K$ (Eq. (1)) for each image *(28, 29)*. See subsection '*CBF and CBV calculations*' for mathematical definitions.

By its nature of operation, SCOS does not require high-speed detectors and can work with commercially available board cameras. Compared to DCS, SCOS can achieve a higher signal-to-noise ratio by leveraging the greater number of speckles collected by the high-pixel count camera *(29, 34, 40, 63)*. On the other hand, SCOS must contend with the relatively high detector noise characteristics of these cameras.

To mitigate this issue, a careful choice of the camera together with a careful calibration of the different sources of noise are necessary *(28, 29, 64)*. For more information about SCOS and interferometric SCOS (iSVS) systems, we recommend the following references *(28–30, 32, 35, 36)*.

### Source-to-detector distance and brain sensitivity

A main component in transcranial optical measurements of CBF and CBV is the source-to-detector (S-D) distance, defined as the distance between the illumination point on the head (source) and the detection point (detector) where the scattered light is collected back, Fig. 1.

As the illuminating light penetrates the scalp layer, it scatters in all directions. A portion of this scattered light penetrates deeper into the skull and brain layers. Some of the light that reaches the brain is scattered backward, traveling back through the brain, skull, and scalp layers. The detector, located at a S-D distance from the source, collects the scattered light. During this process, only a small fraction of the injected light is collected. When the light is coherent or partially coherent *(65, 66)*, such as laser light, the multiple scattering events alter the effective optical path lengths, creating a speckle field. As the laser light interacts with dynamic elements, such as blood cells, it scatters with specific dynamics, resulting in a fluctuating speckle field. This fluctuation can be quantified using the speckle contrast metric.

The depth to which light penetrates the head is intrinsically related to the S-D distance, but the exact relationship between penetration depth and S-D distance remains unclear. While extensive numerical simulations have modeled light penetration into the head, experimental investigations in this domain are still limited. Moreover, individual variations in scalp and skull thickness at different locations on the head complicate generalization from numerical studies. These simulations can provide only a reference sensitivity calculation based on the specific physiological shape used in the models. Prior numerical studies in near-infrared spectroscopy (NIRS) revealed a characteristic banana-shaped spatial sensitivity profile *(42, 44, 67, 68)*. As the S-D distance increases, the 'banana' extends deeper into the head, Fig. 1B, although accessing deeper brain regions becomes more challenging *(42)*. The collected signal was reported to decrease exponentially with depth, presenting a trade-off between improved sensitivity to deeper brain layers and maintaining an adequate signal-to-noise ratio (SNR). Notably, significant gains in brain signal sensitivity have been observed with increases in short S-D distances. Simulations suggest that brain and scalp sensitivity reach parity at approximately 2.5 to 3 cm, with brain sensitivity continuing to increase beyond this range. However, due to the trade-offs of reduced signal strength and spatial resolution, most setups avoid increasing the S-D distance beyond 4 cm.

Building on prior numerical studies and due to lower SNR with fiber-based SCOS, we designed our seven-channel SCOS system with S-D distances ranging from 0.6 cm to 2.6 cm. The specific distances for each channel were as follows: Channel 1: 0.6 cm, Channel 2: 1.0 cm, Channel 3: 1.3 cm, Channel 4: 1.6 cm, Channel 5: 2.0 cm, Channel 6: 2.3 cm, and Channel 7: 2.6 cm.

### Previous indirect verifications that SCOS and others can measure brain blood signal

Studies have indirectly verified the capability of SCOS systems and other optical methods to measure brain blood signal through several approaches:

- A fiber-based SCOS system performed human brain function measurements during a mental subtraction task at S-D = 3.3 cm, observing uprise changes in CBF during the mental task on three human subjects *(29)*.
- The depth sensitivity of interferometric SCOS (iSVS) for measuring CBF was experimentally investigated by tuning the S-D distance from 0.5 cm to 3.2 cm in 0.1 cm increments *(35)*. This study identified the transition point at which CBF in humans and rabbits begins to be detected. Results from three human subjects and two rabbits estimated the S-D threshold distance for detecting CBF to be approximately 1.6 cm for humans and 1.3 cm for rabbits *(35)*. These findings were consistent with comparative MRI and X-ray scans.
- SCOS measurements during breath-holding produced CBF and CBV time traces consistent with transcranial Doppler ultrasonography in a cohort of 23 subjects *(25)*. These findings underscore SCOS's capability to detect brain physiological changes and measure cerebral perfusion. Although an ideal comparison would involve simultaneous SCOS and fMRI scanning, this remains challenging due to the MRI's magnetic environment and the metallic components of the SCOS system. A S-D distance of 3.6 cm was used.
- SCOS measurements during breath-holding in 50 subjects, divided into low- and higher-risk stroke groups, showed differences in CBF and CBV dynamics *(27)*. The low-risk group had a smaller increase in blood flow but a greater increase in blood volume, suggesting that more blood could pass through widened vessels to accommodate better for the breath-holding test. The findings underscore SCOS's capability to assess brain vascular health. A larger cohort is needed to further validate these results. S-D distances of either 3.2 cm or 4.0 cm were used.
- Multi-channel SCOS was used to simultaneously measure CBF and CBV. The brain injury location on one subject was investigated by placing six channels around the head and comparing the results with MRI scans *(41)*. Note that this study involved only one subject. A S-D distance of 3.2 cm was used.
- A bedside SCOS system measured CBF and CBV on 135 subjects for the detection of large vessel occlusion in stroke patients *(31)*.
- Functional interferometric diffuse correlation spectroscopy (also known as interferometric diffusing wave spectroscopy) was employed to measure brain function via CBF signals on four subjects *(43, 69)*. A S-D distance of 3.5 cm was used.
- A scattering phantom experiment demonstrated SCOS's ability to accurately measure flow rates by correlating decorrelation time changes with expected liquid flow in a controlled environment *(28, 35)*.
- A fiber-based SCOS system was employed on 20 mice (10 control, 10 brain injured) to detect traumatic brain injury by measuring changes in CBF *(47)*. A S-D distance of 0.5 cm was used.

While the above studies offer valuable insights suggesting that brain signals can be detected, none have experimentally characterized the sensitivity to scalp and brain blood flow and volume. Although *(43)* explored this sensitivity numerically, the experimental validation remains an open question.

**Experimental Arrangement**

The experimental setup of our aggregated 7-channel SCOS device is illustrated in Fig. 1. Figure 1A presents a 3D visualization of the device positioned on a subject's head, typically placed over the frontal temporal region above the superficial temporal artery branches. Occlusion was achieved by gently applying pressure near the ear bone (Fig. 1A hand diagram). The lower section of Fig. 1A displays a representative CBF time trace from Channel-1 (the closest to the laser source), where a significant dip in the signal during the occlusion period (8 to 17 seconds) confirms a reduction in the scalp blood flow.

Figure 1B provides top and lateral views of the 7-channel device, highlighting the detection channels at different S-D distances designed to sense the scalp, skull, and brain layers. The device comprised one multimode optical fiber laser source and seven detection multimode fibers, all housed within a 3D-printed mount. The mount was fabricated using a resin-based printer [Anycubic Photon Mono X 6K] with black resin to enhance light absorption and minimize back reflections and stray light. Post-printing, the components were fully cured under UV light to ensure safe contact.

On one side of the mount, a 600 μm core diameter multimode fiber [Thorlabs M29L01] delivered laser light to the subject's head. The laser was a thermally stabilized 830 nm laser [Crystalaser DL830-300-SO] with a long coherence length (>5 m), high beam quality factor ($M^2<1.3$), single-longitudinal mode, and built in optical isolator. Although the laser source could provide up to 300 mW, the output power was limited to 60 mW using a waveplate and a polarized beam splitter before coupling the laser light into the optical fiber (details in Supplementary Material Fig. S1). The fiber was positioned 8 mm from the skin and was oriented with an angle of 10° from perpendicular illuminance, producing a 5 mm illumination spot diameter, similarly than in *(27, 28, 35, 41)*. This setup ensured the laser intensity remained within the American National Standards Institute (ANSI) safety limit for skin exposure to an 830 nm laser beam (3.63 mW/mm²) *(45)*.

On the opposite side, seven 600 μm core diameter multimode fibers were positioned at varying S-D distances from the illumination fiber, each representing a separate SCOS channel. These fibers were in direct contact with the skin with perpendicular orientation to maximize light collection efficiency. The seven fibers were combined into a single fiber bundle using a 7-to-1 fan-out bundle fiber [Thorlabs BF76HS01]. The output of the bundled fiber was imaged onto a scientific CMOS camera [Excelitas pco.edge 5.5] featuring a 6.5 × 6.5 μm pixel size and 16-bit pixel depth dynamic range. The large dynamic range is essential for this experiment, as it allows simultaneous recording of signals with vastly different intensities—where short S-D distance channels generate high signal intensities, while long S-D distance channels produce much lower signal intensities, often approaching or falling below the dark noise level. The camera operated with an exposure time of 8 ms and was recording at 65 FPS. This setup resulted in a speckle density of 0.1 speckle per pixel, corresponding to a one-dimensional speckle-to-pixel length ratio of s/p = 3 *(28, 40, 63)*. The collected light from the seven detecting fibers was therefore imaged onto the camera simultaneously. To prevent cross-channel light leakage in the recorded camera image, a telescope imaging system was meticulously aligned to magnify and focus on the end plane of the fiber bundle, see Supplementary Material Fig. S1 for details.

The bottom surface of the mount was partially angled to accommodate the curvature of the head, see Supplementary Material Fig. S2. To minimize noise from fibers' movement, the fibers were secured to the mount using clips and glue. The mount had four slots along the sides to strap the device on the head using Velcro straps tight enough to avoid device sliding but to no restrict scalp blood flow due to the pressure of the mount onto the scalp blood vessels, which can alter our results.

For each subject, the device's location on the head was adjusted to align with the region where the superficial temporal artery spreads, ensuring that the occlusion effects were clearly observed (see Supplementary Material Fig. S4 for more inputs). For some subjects, part of the device rested on areas with hair, though efforts were made to minimize this. The device orientation and location were also adjusted until the signal intensity across of at least five channels largely exceeded the camera dark noise level. Variations in scalp and skull thickness among individuals led to differences in ideal S-D distances *(35)*. Some subjects displayed CBF and CBV signals across all seven channels, while others exhibited signals across five channels. No subjects exhibited signals for fewer than five channels. Once secured, a 24-second benchmark trial was conducted to confirm CBF and CBV signal quality for Channels 1 to 5 and to ensure no movement artifacts were present. The temporal occlusion and control experiments were conducted only after ensuring proper device positioning and signal quality.

**CBF and CBV calculations**

In this section, we briefly describe the methods for extracting CBF and CBV from speckle camera images in SCOS. From the collected camera images, we segmented the camera image into seven parts, one for each channel (see Supplementary Material Figure S3). From each channel, we calculated the speckle contrast K as:

$$K_i^2(I_i(t)) = \frac{\sigma^2(I_i(t))}{\mu^2(I_i(t))}. \tag{1}$$

Where $\sigma^2(I(t))$ is the variance of the normalized image $I$ of channel $i$ at time $t$ and $\mu(I(t))$ its mean. The various sources of noise were accounted as *(28, 29, 70, 71)*:

$$K_{adjusted}^2(t) = K_{raw}^2(t) - K_{shot}^2(t) - K_{quant}^2(t) - K_{cam}^2(t) \tag{2}$$

with $K_{shot}^2$ accounting for variance contributions from the shot noise, $K_{quant}^2$ for the variance inherited from quantization, and $K_{cam}^2$ for the variance contributions of the camera's readout noise and dark noise. See Ref*(28)* for more details about the speckle contrast calculations and calibration processes. The blood flow index (BFI) was related to $K_{adjusted}^2$ by:

$$BFI(t) = \frac{1}{K_{adjusted}^2(t)}. \tag{4}$$

The blood volume was extracted from the camera images by calculating the change of blood volume index ($BVI$) over the baseline based on logarithmic definition derived from modified Beer-Lambert law, as defined in Ref. *(29)*:

$$BVI(t) = \log_{10}(\frac{I_0}{\mu(I(t))}), \tag{5}$$

where $I_0$ is the intensity at baseline, calculated as the mean of the average intensity of the entire image acquired during the baseline period (first 7 seconds). The mean $\mu(I(t))$ was calculated as the average spatial intensity of the entire image $I$ acquired at time $t$, and can be expressed in camera grayscale units. In cases without significant or acute volume change, the logarithmic equation shown in Eq. (5) can be linearly approximated, and blood volume index can be evaluated with the equation shown as *(25, 27)*:

$$BVI(t) \approx \frac{2I_0 - \mu(I(t))}{I_0}. \tag{6}$$

**Occlusion of the superficial temporal artery and its effects**

The carotid artery is a major blood vessel that supplies blood to the head. It divides into two major branches: the internal carotid artery, which supplies blood to the brain, and the external carotid artery, which supplies blood to the scalp, skull, and other superficial structures. The external carotid artery begins at the carotid bifurcation (near the Adam's apple) and travels up toward the ear on either side. Near the ear, it divides into two terminal branches: the maxillary artery and the superficial temporal artery.

The superficial temporal artery, originating below the ear and running vertically between the cheekbone and the ear, supplies blood to the scalp and face but does not contribute to the brain's blood supply *(72)*. The ear bone, positioned behind the temporal artery, serves as a stable and rigid support platform for performing occlusion by applying gentle pressure. When performed temporarily (i.e., for less than a few minutes), this occlusion technique is entirely safe on subjects with no history of superficial temporal artery pathology. In addition, the superficial temporal artery can be easily located by palpating for a pulse near the ear bone. A control experiment, where similar pressure is applied near the top of the ear without affecting blood flow, can further validate the reliability and precision of this approach.

These factors make the superficial temporal artery an excellent candidate for isolating scalp blood flow from cerebral circulation though a temporary occlusion: 1) it is a safe and minimally invasive procedure, 2) it only blocks blood flow to the scalp and skull and does not impact cerebral circulation, 3) the temporal artery is easy located and is simple to occlude by applying gentle pressure, 4) it is a repeatable procedure across different subjects, and 5) it has a straightforward control experiment equivalent.

**Human Research Study**

A total of 20 healthy adult subjects participated in our experiments, comprising 9 females and 11 males with ages ranging from 20 to 77 years. This diverse cohort ensured a balanced representation across sex and age groups in the results. No subjects were excluded from the analysis following quality control of the raw data, which ensured the presence of a significant dip in blood dynamics during superficial temporal artery occlusion. This underscores the high repeatability and robustness of our experimental protocol.

Before the experiments, each participant completed a questionnaire (shown in the supplementary of *(27)*), and we measured their blood pressure. Informed consent was obtained from each participant beforehand. The human research protocol for this study received approval from the Caltech Committee for the Protection of Human Subjects and the Institutional Review Board (IRB) under protocol IR21-1074. The total illumination power was within the American National Standards Institute (ANSI) laser safety standards for maximum skin exposure of an 830 nm laser beam *(45)*. To simplify the experiment and system implementation, measurements were taken on the temple area with minimal or no hair. For all the 20 subjects, CBF and CBV were recorded for 24 seconds, with the temporal occlusion (or control press) applied from the 8$^{th}$ to the 17$^{th}$ second.

## Acknowledgments

We thank the Caltech Summer Undergraduate Research Fellowship (SURF) program and the Chung Ip Wing-Wah Memorial Foundation for their support.

**Funding:**
National Institutes of Health — Award No. 5R21EY033086-02
Caltech Center for Sensing to Intelligence (S2i) — Award No. 13630012
USC Neurorestoration Center
2024 SPIE-Franz Hillenkamp Postdoctoral Fellowship

# Supplementary Material for:
# Assessing Sensitivity of Brain-to-Scalp Blood Flows in Laser Speckle Imaging by Occluding the Temporal Artery

**Abstract**
This Supplementary Material contains six section and six supplementary figures. In this material, we provide comprehensive details on the experimental setup, methodologies, analyses, and supporting results that substantiate the main findings of our study. Section S1 outlines the experimental arrangement, including the optical configuration for delivering light into the laser source fiber and imaging collected light from the seven-channel SCOS system. Section S2 describes the design and configuration of the optical fiber device, focusing on fiber mounting and the optimized head-mount design for human use. Section S3 details the processing workflow for simultaneous imaging of the light collected by the seven detector fibers and the extraction of blood flow index (BFI) values. In Section S4, we show the effect of a large blood vessel crossing channels with long S-D distances and its implications for temporal occlusion signal sensitivity. Section S5 presents the results for the ratio of < BFI > across the 20 study subjects, highlighting inter-subject variability. Finally, Section S6 provides an example of correlation matrices for BFI and BVI pulse shape datasets. Then, the section shows the combined results for BFI and BVI correlations from 20 subjects. Collectively, these sections provide a comprehensive exploration of the experimental methods utilized in our research.

## Section S1 – Detailed Experimental Arrangement

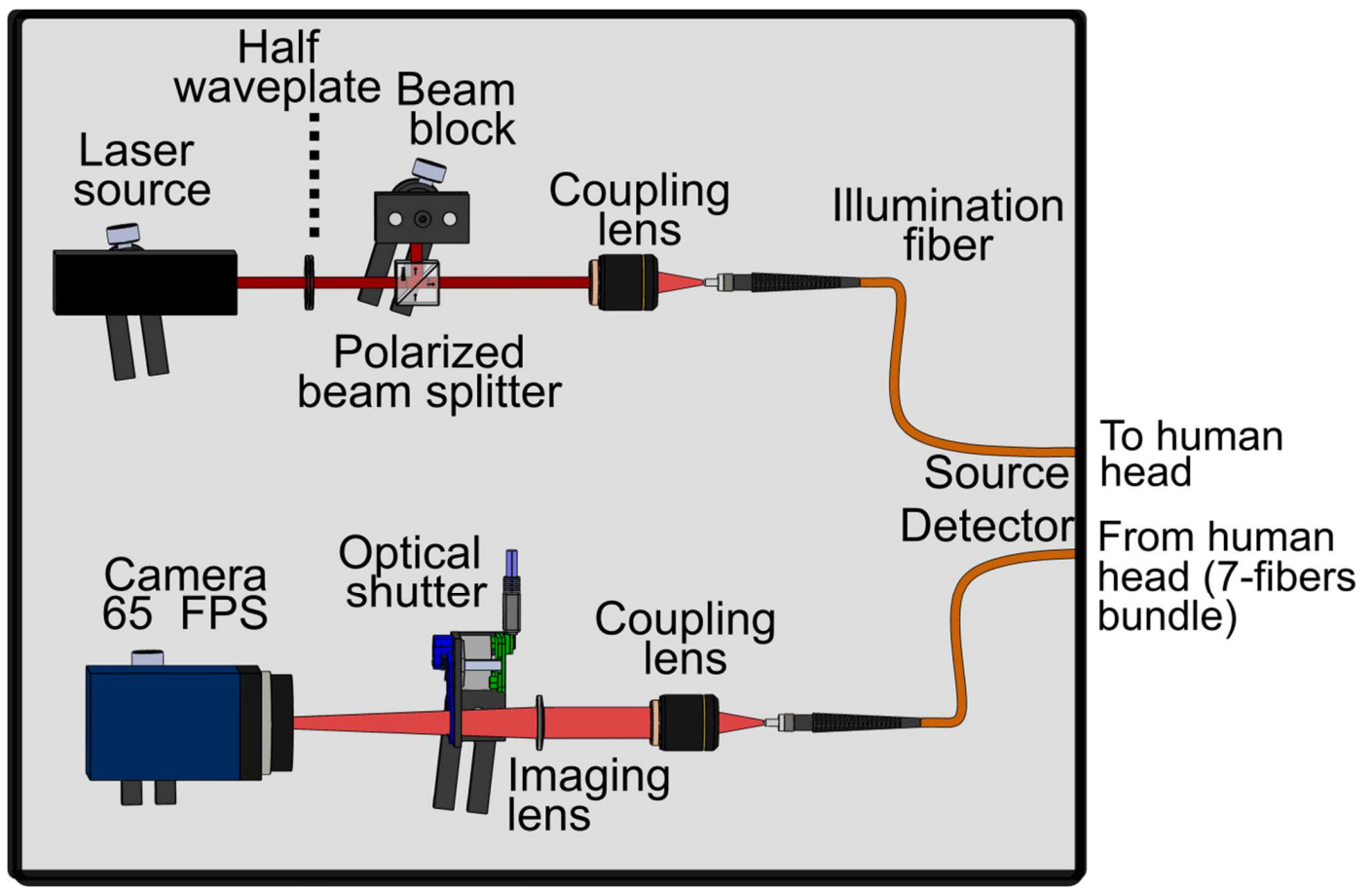

**Fig. S1:** Detailed experimental arrangement designed to deliver light into the illumination source and capture the light collected from the seven detecting fibers into a single camera.

# Section S2 – Optical Fibers Device Configuration

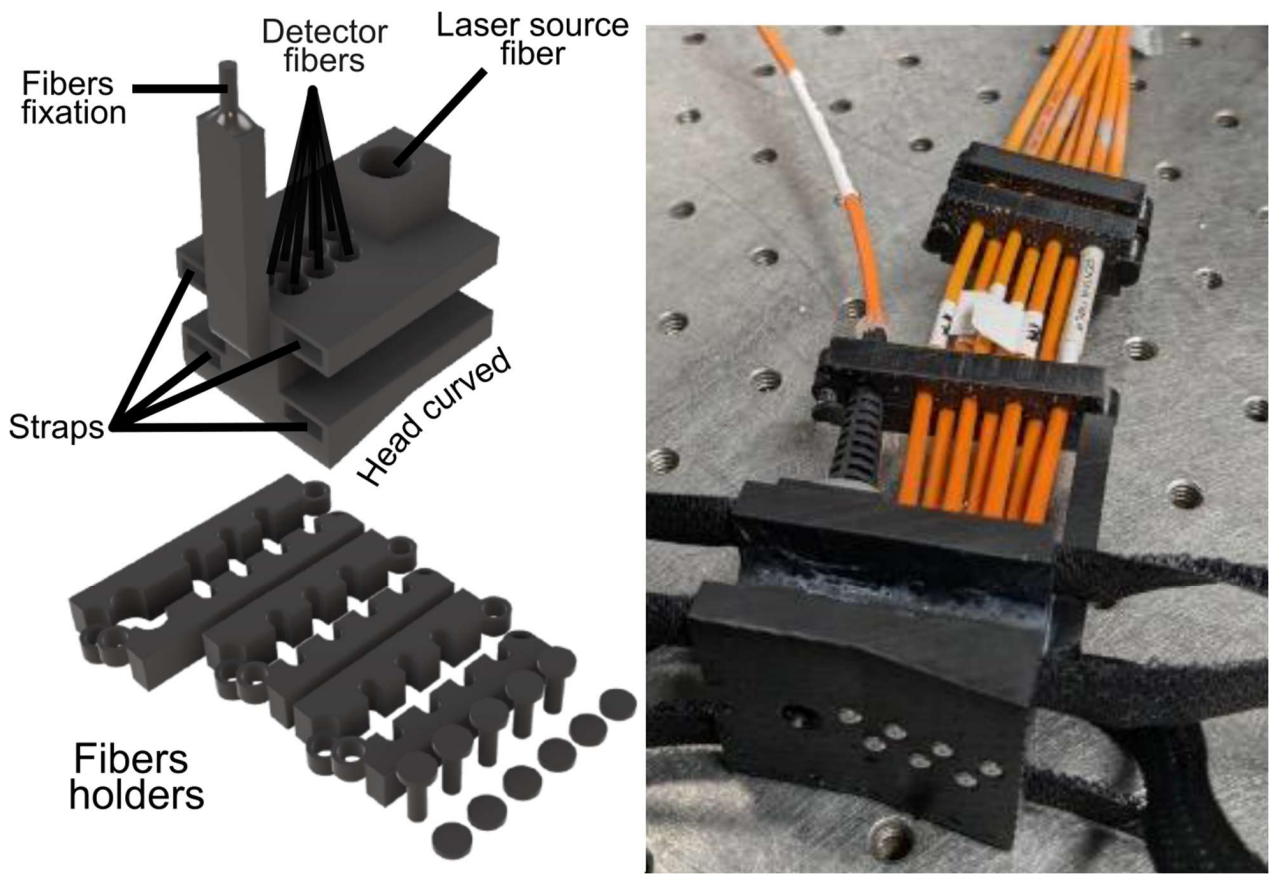

**Fig. S2:** A detailed view of the device mount designed to secure the optical fibers within the 3D-printed structure and strap it around the head. The optical fibers are held in place using custom-made fiber holders and glue. Four slots along the edges of the main mount accommodate straps for securing the device to the head. The bottom surface of the main mount is angled to conform more closely to the curvature of the head, ensuring a snug and stable fit. The fiber holders also act as clips to bundle the collection fibers together, reducing fibers movement and enhancing stability. Right photo: The fully assembled and functional device, showcasing the completed design.

# Section S3 – Detector Fibers Simultaneous Imaging and BFI Calculations

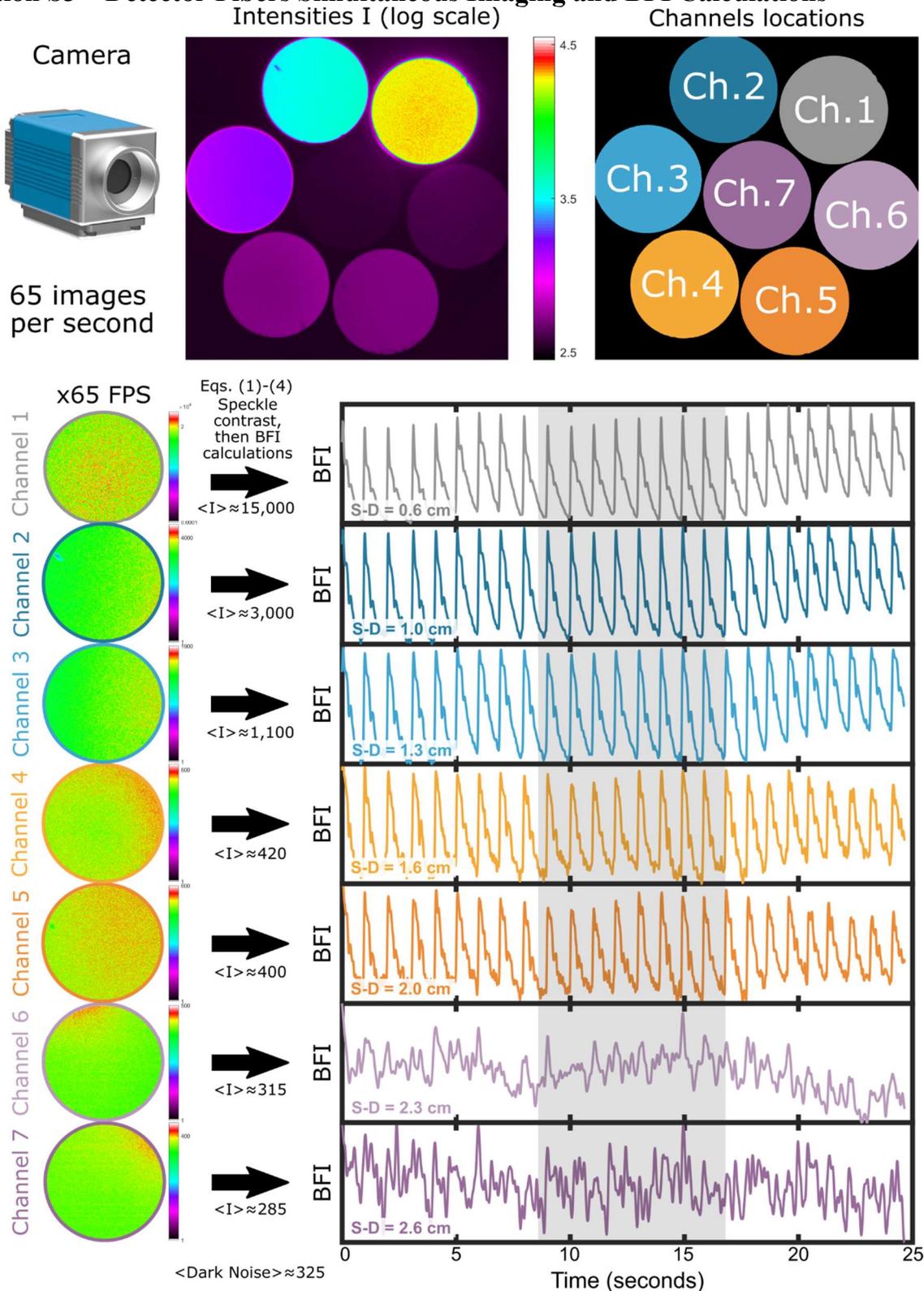

**Fig. S3:** Processing Chart: from camera images capturing signals collected from the seven fibers to speckle contrast calculations to extract the Blood Flow Index (BFI) for each channel.

**Section S4 – Case of a Blood Vessel Crossing Channels at Long S-D Distances**

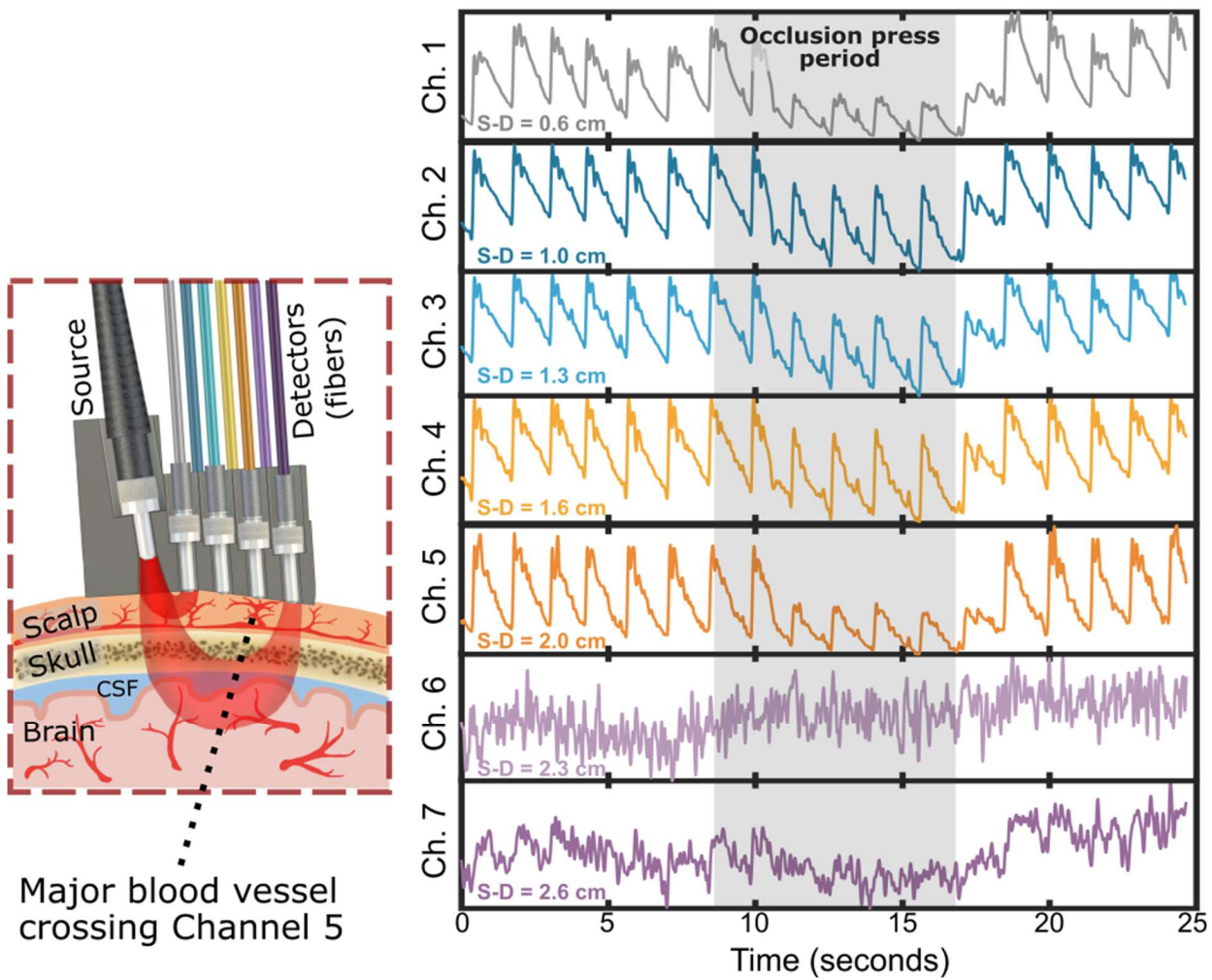

**Fig. S4:** A specific case of a major scalp blood vessel crossing a single channel (here Ch. 3 and Ch. 4), which altered the temporal occlssusion results. In this instance, the prominence of the scalp blood vessel was evident by the naked eye. This highlights the critical importance of ensuring a uniform distribution of scalp blood vessels across all channels to maintain consistency in measurements.

# Section S5 – Ratio of < BFI > Across the 20 Subjects

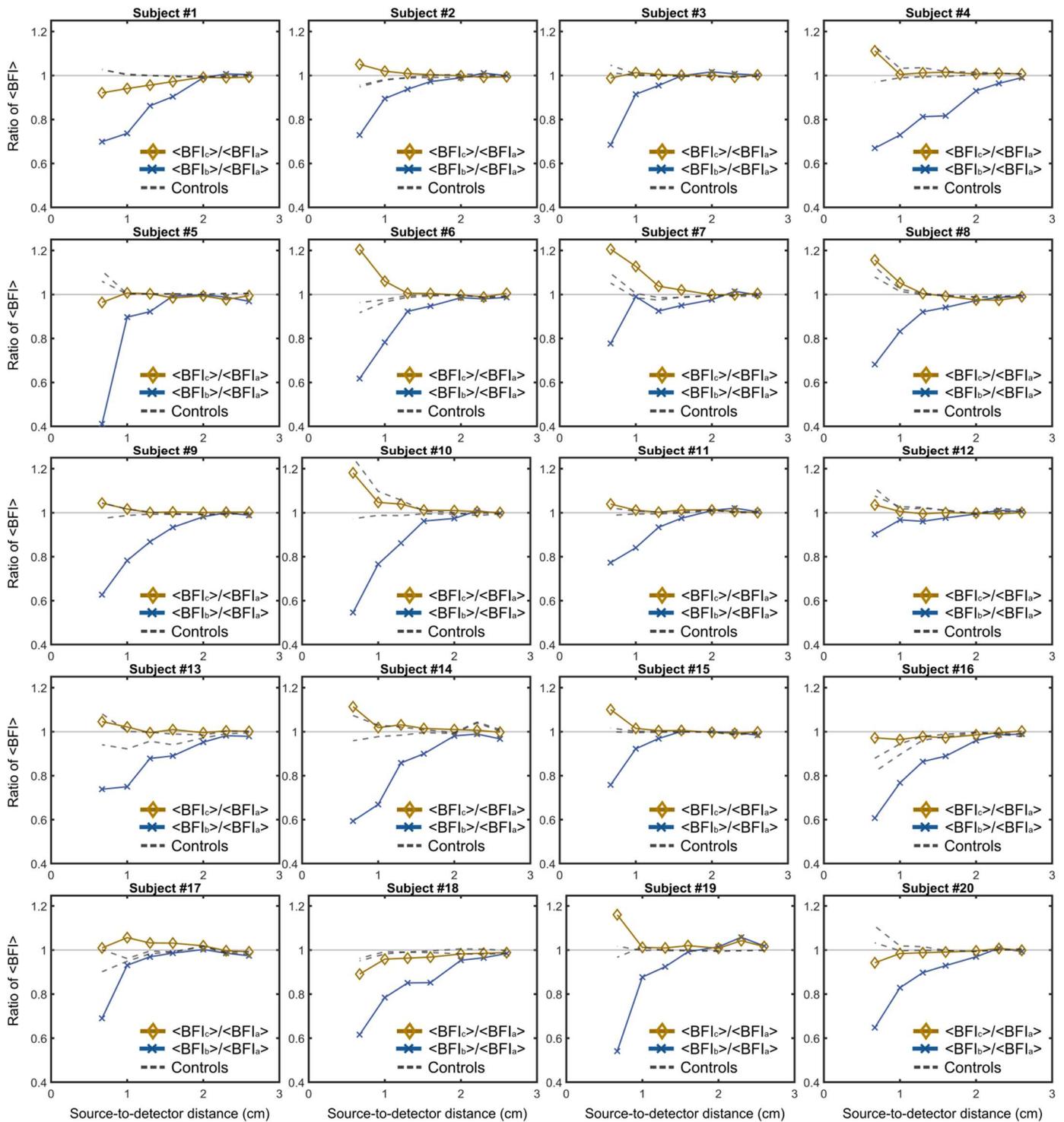

**Fig. S5:** Ratio of < BFI > for each of the 20 subjects shown in Fig. 3B.

# Section S6 – Correlation matrix in BFI and BVI datasets

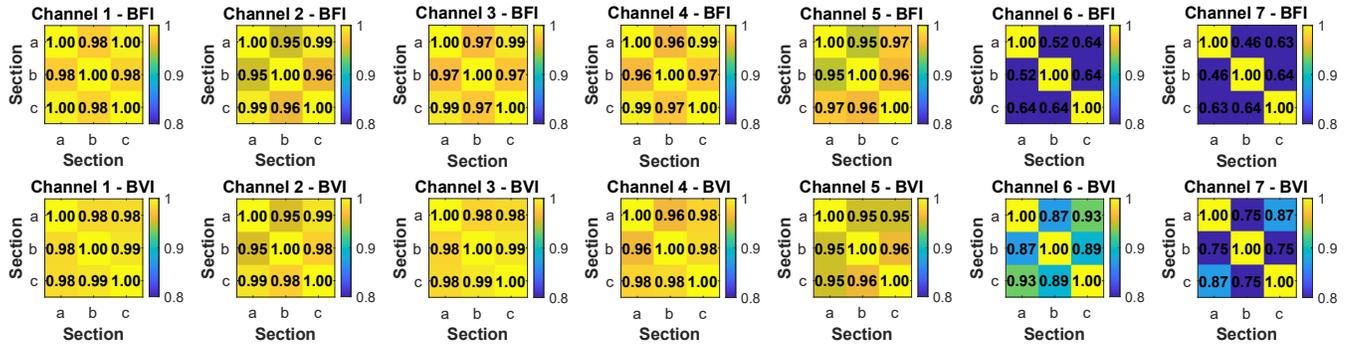

**Fig. S6:** An example of Pearson correlation matrices for BFI (blood flow index) and BVI (blood volume index) from subject 1. We segmented and normalized the individual cardiac periods of BFI and BVI from each section, then calculated the average pulse shapes. Using this method, we constructed confusion matrices for each channel. In this particular example, BFI exhibits relatively low signal-to-noise ratio (SNR) for channels 6 and 7, and BVI shows relatively low SNR for channel 7. The section labels are as follows: section a – pre-occlusion, section b – during occlusion, and section c – post-occlusion.

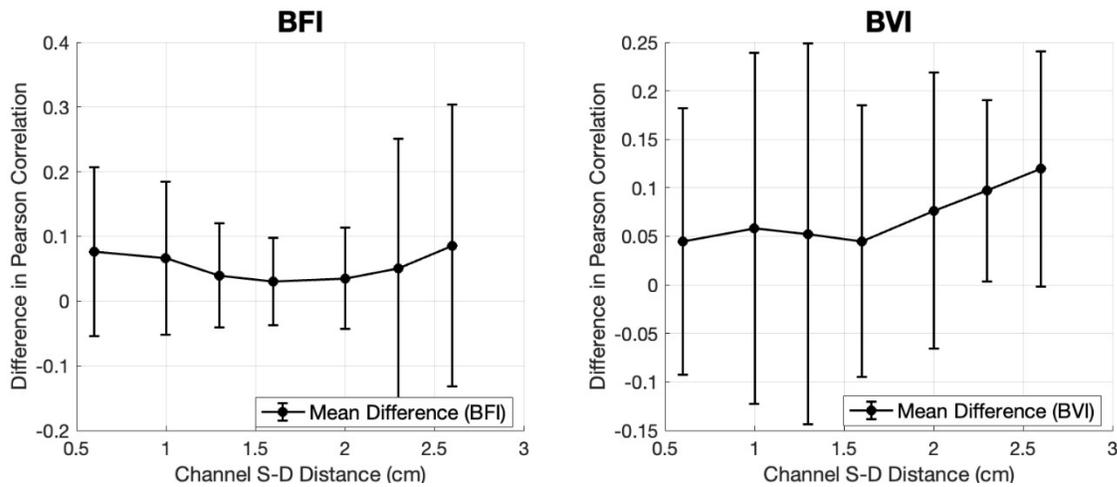

**Fig. S7:** The difference in Pearson correlations for BFI and BVI. This difference is calculated by subtracting the correlation between the post-occlusion and pre-occlusion pulses ($\rho_{ac}$) by the correlation between the during-occlusion and pre-occlusion pulses ($\rho_{ab}$). The error bars are notably larger for BFI in the last two channels due to the relatively lower SNR.

Our hypothesis posits that the BFI periodic shape during occlusion differs from the pre-occlusion shape, with the differences being most pronounced in channels with smaller source-detector (S-D) separations (notably channels 1 and 2) due to stronger scalp blood flow influence. Conversely, we hypothesize that the post-occlusion shape will closely resemble the pre-occlusion shape. This expected difference in Pearson correlation, defined as $\rho_{diff} = \rho_{ac} - \rho_{ab}$, is visualized in Fig. S7.

While we observed a decreasing trend in $\rho_{diff}$ as the S-D distance increased in multiple subjects, the final results from the 20-subject dataset are inconclusive. Consequently, we are refraining from drawing definitive conclusions at this stage. As shown in Fig. S7, the error bars are larger for smaller S-D separations, likely due to variations in occlusion pressure and slight differences in surface blood vessel positioning across subjects. The effect of these variations diminish with moderate S-D distances (>1 cm, corresponding to channels 3–5), where scalp sensitivity decreases and brain sensitivity increases. In this range, both $\rho_{diff}$ and error bars decrease. The trend reverses at larger S-D separations, where SNR decreases. However, $\rho_{diff}$ consistently remains positive, suggesting the occlusion will change the BFI cardiac pulse shape.

We also included BVI period correlations in Fig. S7. The calculation of the BVI pearson correlation differences are the same as that of BFI, but we do not observe the same trend as BFI. The difference in pearson correlations for BVI does not converge at channel 3 to channel 5 as shown in BFI. This may also be due to the lower brain sensitivity of BVI relative to BFI at the same S-D distance.